# Modal Decomposition of Turbulent Supersonic Cavity

Rahul Kumar Soni, Nitish Arya, and Ashoke De*
Computational Propulsion Lab, Department of Aerospace Engineering, Indian Institute of Technology Kanpur, India

*Corresponding author: Ph:+91-5122597863, Fax: +91-5122597561
E-mail: ashoke@iitk.ac.in

**Abstract:**

Self-sustained oscillation in Mach 3 supersonic cavity with a length-to-depth ratio of 3 is investigated using wall modeled Large Eddy Simulation (LES) methodology for $Re_D = 3.39 \times 10^5$. The unsteady data obtained through computation is utilized to investigate the spatial and temporal evolution of the flow field, especially second invariant of velocity tensor; while the phase averaged data is analyzed over a feedback cycle to study the spatial structures. This analysis is accompanied by the Proper Orthogonal Decomposition (POD) data, which reveals the presence of discrete vortices along the shear layer. The POD analysis is performed in both the spanwise and streamwise planes to extract the coherence in flow structures. Finally, Dynamic Mode Decomposition (DMD) is performed on the data sequence to get the dynamic information and deeper insight into the self-sustained mechanism.

## 1. Introduction

As we know that the supersonic cavity flows have been widely studied both experimentally and numerically [1-4] because of its application to various practical engineering problems. Although geometrically simple, the flow physics is quite challenging and complex. First insight into the rectangular cavity was provided by Rossiter [5]; where a semi-empirical formula was proposed to predict the resonant frequencies. Also, the mechanism of a feedback cycle was explained [6-8] through the sequence of events, such as: (1) excitation of shear layer at the leading edge due to acoustic resonance leading to vortical shedding, (2) convection and growth of the vortical structures further downstream ultimately impacting the trailing edge, (3) generation of upstream travelling compression waves at the trailing edge, and (4) upstream propagation of these compression waves exciting the shear layer. Mechanism of flow induced pressure oscillations in cavities had been reported in published literature [9-12]. Heller and Bliss [10, 11] modified the Rossiter's formula and further classified four types of waves: Type 1: upstream propagation of waves above the cavity, due to the disturbances traveling upstream within the cavity, Type 2: formation of compression and expansion waves at the leading edge due to shear layer oscillation, Type 3: the quasi-steady external bow shocks at the trailing edge due to reattachment of flow, and Type 4: the weak compression waves near trailing edge. Zhang *et al*. [13] were the first one to observe the fifth type of wave propagating upstream as an acoustic wave or perturbation. Arunajatesan *et al*. [14] simulated the flow over a cavity using Large Eddy Simulation as well as hybrid RANS/LES solver. The hybrid model predictions were in good accordance with the measured pressure values. A computational investigation for the supersonic flow (M=1.5) over a 3-Dimensional cavity was done by Rizzetta [15]. The frequency spectra from this computational analysis agreed well with the experimental data. It was noted that while the fundamental behavior of the problem was 2-Dimensional, the presence of a vortex evolving at the side wall of the front wall of the cavity produced noticeable 3-Dimensional effects. In a subsequent study, Zhuang *et al*. [16] observed shock waves generation in shear layer due to convection of large scale structures. At high Reynolds number, the flow above a rectangular open cavity is relatively undisturbed and the prominent unsteady flow feature is the convection of the vortices in the shear layer [17]. The dynamics of such a flow can be studied by observing the mechanism of the generation and propagation of the shear layer vortices. But, any high Reynolds number flow is associated with the huge separation of scales and isolating the different types of vortices present in the unsteady shear layer is a humongous task. Therefore, the Proper Orthogonal Decomposition (POD) and the Dynamic Mode Decomposition (DMD) are employed for the present study.

The proper Orthogonal Decomposition (POD), also known as Karhunen-Loeve Expansion or Principle Component Analysis was first introduced by Lumley [18] in the context of fluid dynamics to understand the turbulent flows by identifying the coherence in flow structures. The large scale structures, also called the coherent structures, are primarily responsible for most of the transport occurring in any turbulent flow. Thus, the study of turbulent flows which contain a wide range of scales and infinite degree of freedom reduces to the study of dynamics of the coherent structures. POD involves the decomposition of the flow field in time and space and the Eigen modes are computed by collecting sufficient number of snapshots separated properly in time as temporal separation plays vital role in the decomposition process. The decomposition is performed in such a manner that only few basis functions are able to represent the most of the energy. Basically, this allows only first few modes to represent the flow field accurately. Over the years, the POD has emerged as a powerful tool for the extraction of dominant structures in turbulent flows. The POD has been widely used to construct a low-dimensional model to study the dynamics of the unsteady flow field. There exist various types of POD techniques of which, direct method and method of snapshots are used widely. In the present investigation, the method of snapshots proposed by Lawrence [19] is utilized and the formulation of energy based POD is similar to [20-23].

The POD modes obtained from any system are based on the energy content of the loss of phase information which makes POD unsuited for capturing any dynamic information of the system. Thus, any small perturbation leading to large-scale instability will not be captured by POD. However, the resulting flow features with high energy content will be reflected in the POD modes. A system comprising of a local instability in any finite region is capable of exhibiting self-sustained oscillations at a particular frequency. The flow over an open cavity is one such example with the deflection of the shear layer being the local instability. Thus, to extract dynamical features of a system, Dynamic Mode Decomposition (DMD), a tool based on Koopman analysis [24] and introduced by Schmid [25], is employed. The basis of the algorithm is the extraction of low dimensional subspace after fitting a high degree polynomial to the original data sequence without any prior knowledge of the process by which the data had initially been generated. Thus, despite employing an infinite dimensional linear Koopman operator [24, 26], DMD can be used for any data which stems from a linear or a non-linear process. The eigenvalues and eigenvectors of the low-dimensional subspace capture the principal dynamics of the flow. Basically, DMD attempts to represent a data sequence by orthogonalizing it in time, while POD attempts a decomposition based on orthogonality in space. Furthermore, the DMD is directly applied to the data, while a POD analysis processes second-order statistics of the data.

The POD and DMD are powerful tools to study any turbulent flow associated with an enormous number of scales (length and time) with an infinite degree of freedom. They convert the study of the turbulent flow into the analysis of few modes corresponding to coherence in space and time. The present flow geometry involves various interactions between the acoustic and the hydrodynamic modes which lead to the generation of different types of vortices. The spatial and temporal evolution of these structures is key to understand the overall flowfield. Hence, in the present work, both the modal decomposition techniques are employed to study the generation and the propagation of the vortices along the shear layer and the feedback loop generated within the cavity which is the consequence of the convection of these vortices. To the authors' best knowledge POD and DMD analysis of a supersonic cavity flow has not been reported widely except recent work by Zhang *et al*. [27]. Their work involves reduced order modelling for supersonic flow over a cavity using POD and Galerkin projection, whereas the present research focuses on the study of the hydrodynamic and acoustic modes due to the convection of discrete vortices along the shear layer. To be precise the major philosophical difference between our study and that of the study of by Zhang *et al*. [27] lies in the fact that they propose a new norm for the reduced order modelling of the supersonic flow. However, in the present investigation, we are only interested in the modal decomposition and not the reduced order modelling, by this we mean that these decomposition tools are utilized to characterize the structures generated due to the hydrodynamic and acoustic interactions leading to the self-sustained oscillation.

## 2. Numerical details

The filtered governing equations for the conservation of mass, momentum, and energy are solved and recast as:

Continuity equation:

$$\frac{\partial}{\partial t}(\bar{\rho}) + \frac{\partial}{\partial x_i}(\bar{\rho}\tilde{u}_i) = 0 \qquad (1)$$

Momentum equation:

$$\frac{\partial}{\partial t}(\bar{\rho}\tilde{u}_i) + \frac{\partial}{\partial x_j}(\bar{\rho}\tilde{u}_i\tilde{u}_j) = -\frac{\partial}{\partial x_i}(\bar{p}) + \frac{\partial}{\partial x_j}\left[(\mu+\mu_t)\frac{\partial \tilde{u}_i}{\partial x_j}\right] \qquad (2)$$

Energy equation:

$$\frac{\partial}{\partial t}(\bar{\rho}\tilde{E}) + \frac{\partial}{\partial x_i}(\bar{\rho}\tilde{u}_i\tilde{E}) = -\frac{\partial}{\partial x_j}\left[\tilde{u}_j\left(-\tilde{p}I + \mu\frac{\partial \tilde{u}_i}{\partial x_j}\right)\right] + \frac{\partial}{\partial x_i}\left[\left(k + \frac{\mu_t C_p}{\Pr_t}\right)\frac{\partial \tilde{T}}{\partial x_i}\right] \qquad (3)$$

Equation of State

$$\tilde{p} = \bar{\rho}R\tilde{T} \qquad (4)$$

Where ρ is the density, $u_i$ is the velocity vector, p is the pressure, $E = e + u_i^2/2$ is the total energy, where $e = h - p/\rho$ is the internal energy and h is enthalpy. The fluid properties μ and k are respectively the viscosity, and the thermal conductivity, while $\mu_t$ and $\Pr_t$ are the turbulent eddy viscosity, and the turbulent Prandtl number, respectively. The only unclosed terms in the above set of equations are the subgrid stresses (terms involving $\mu_t$) which are modeled with the help of turbulent eddy viscosity formulation.

The (-) quantities in the above equations are the Favre averaged quantities and the (~) quantities are the ones that are obtained after the application of the Filter function. The application of the Filter in the momentum equation produces SGS (Subgrid Scale) Stresses which are modeled using an eddy viscosity assumption. The turbulent eddy viscosity is given by the relation-

$$\mu_t = (C_s \bar{\Delta})^2 |\bar{S}_{ij}| \tag{5}$$

Where $C_s$ is the Smagorinsky constant $\bar{\Delta}$ is the filter width and $|\bar{S}_{ij}| = (2\bar{S}_{ij}\bar{S}_{ij})^{\frac{1}{2}}$ is the strain rate magnitude. The Smagorinsky constant is determined apriori and is kept constant throughout the whole domain and computational time. Yoshizawa [28] proposed an eddy viscosity model which uses the Smagorinsky model to account for the anisotropic part of the SGS stress tensor while the SGS energy was modeled separately as presented in the equations

$$\tau_{ij} - \frac{\delta_{ij}}{3}\tau_{kk} = -C_s^2 2\bar{\Delta}^2 \bar{\rho}\left(\tilde{S}_{ij} - \frac{\delta_{ij}}{3}\tilde{S}_{kk}\right) = C_s^2 \alpha_{ij} \tag{6}$$

$$\tau_{kk} = C_I 2\bar{\rho}\bar{\Delta}^2 \left|\tilde{S}_{ij}\right| \tag{7}$$

The constant value of the Smagorinsky constant $C_s$ hinders the accurate prediction of the fields in massively separated flow regions which is overcome by using a dynamic procedure for the determination of the Smagorinsky constant [29]. This procedure involves an application of a test filter $\hat{\Delta}$ which is roughly two times the size of the filter $\bar{\Delta}$. The SGS stresses $\tau_{ij} = \overline{u_i u_j} - \overline{u_i}\,\overline{u_j}$ are related to the resolved turbulent stresses $L_{ij} = \widehat{\widetilde{u_i}\widetilde{u_j}} - \widehat{\widetilde{u_i}}\widehat{\widetilde{u_j}}$ and the subtest stresses $T_{ij} = \widehat{\overline{u_i u_j}} - \widehat{\widetilde{u_i}}\widehat{\widetilde{u_j}}$ via Germano identity [28]. A dynamic procedure was applied by Moin *et al*. [27] for the two constant model proposed by Yoshizawa [27]. For the present study, this dynamic model has been used where the constants $C_s$ and $C_I$ were calculated as follows-

$$C = C_S^2 = \frac{\langle L_{ij} M_{ij} \rangle}{\langle M_{kl} M_{kl} \rangle} \qquad \text{and} \qquad C_I = \frac{\langle L_{kk} \rangle}{\langle \beta - \alpha \rangle} \tag{8}$$

$$\left. \begin{array}{l} \beta_{ij} = -2\hat{\Delta}^2 \hat{\bar{\rho}} \left|\breve{\bar{S}}\right| \left(\breve{\bar{S}}_{ij} - \delta_{ij} \breve{\bar{S}}_{kk}/3\right) \\[2mm] M_{ij} = \beta_{ij} - \hat{\alpha}_{ij} \\[2mm] \beta = 2\hat{\Delta}^2 \hat{\bar{\rho}} \left|\breve{\bar{S}}\right|^2 \end{array} \right\} \tag{9}$$

In order to reduce the computational cost of the simulation, the present LES grid does not resolve the wall boundary layer. Instead, a wall model is used to generate a smooth profile from the wall up to the first grid point. The modelling of the boundary layer is achieved through a wall model which uses Spalding's law of the wall to create a smooth profile for $\mu_{sgs}$ from the wall until the first grid point. The Spalding's law of the wall is given by

$$y^+ = u^+ + 0.1108(e^{0.4u^+} - 1 - 0.4u^+) \tag{10}$$

Where,

$$u_\tau = \sqrt{\tau_w/\rho} \ , \ y^+ = yu_\tau/v \ , \ u^+ = u/u_\tau \tag{11}$$

From Spalding's law, a smooth profile for the turbulent eddy viscosity $\mu_t$ is created from the wall up to the first grid point.

$$\frac{\mu_t}{\mu_{eff}} = 1/[1 + \frac{1}{0.04432}\{e^{0.4u^+} - 1 - 0.4u^+ - \frac{(0.4u^+)^2}{2!}\}] \tag{12}$$

The performance assessment of the present wall model in predicting the velocity profile is done by Soni *et al*. [31] by comparing the results of LES simulation with the DNS results of [32, 33]. It has been shown by [31] that the values predicted by the wall model for y⁺ less than 100 are in excellent agreement with the DNS results.

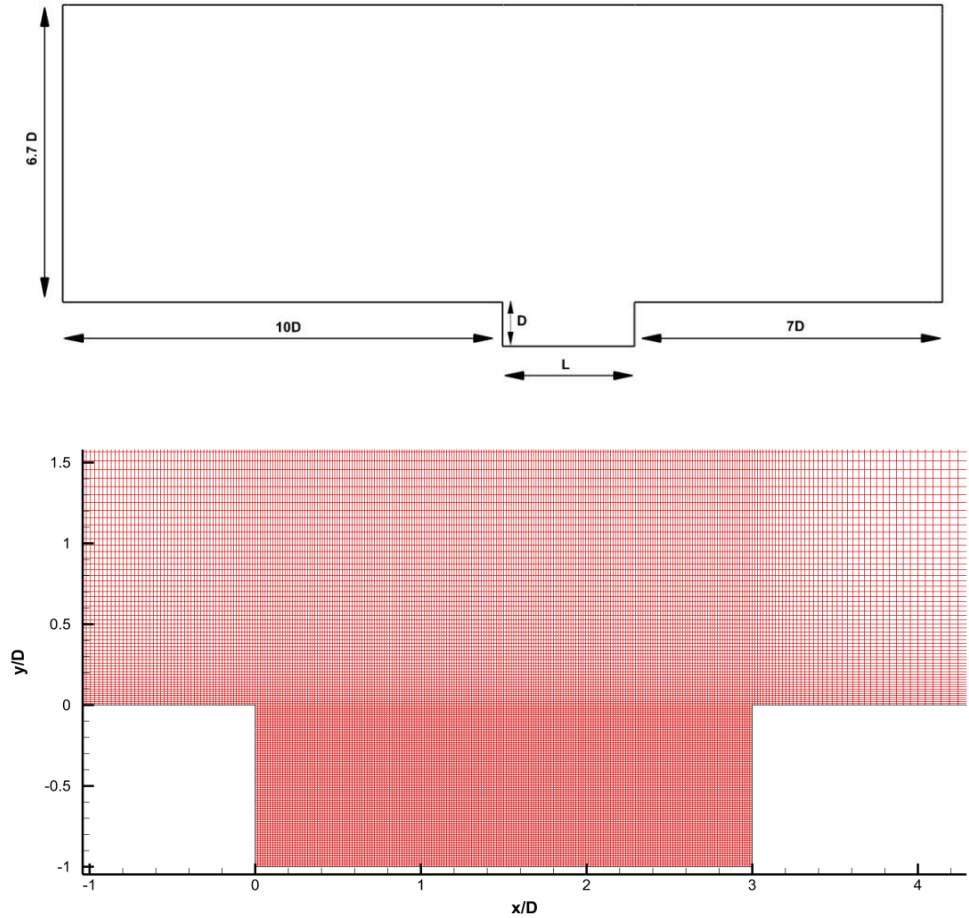

Figure 1: Computational domain (xy plane) and grid distribution

## 3. Computational details

The computational domain consists of two regions: such as internal (cavity) region and outer (channel) region as shown in Figure 1. The length from domain inlet to the cavity leading edge is 10D and from the trailing edge to domain outlet, it is 7D to allow the reattachment shock to exit the system without reflecting. In the outer region, the length of the wall normal direction is 6.7D and in the spanwise direction, it is 2D. The cavity has L/D ratio of 3 where D = 8.9 mm. The domain is discretized through the multi-block approach to maintaining optimum grid size while retaining the resolution in the region of interest. Three sets of grids are generated namely Grid 1, Grid 2 and Grid 3. For Grid 1, the outer region has node distribution of 250 ×85×40, whereas in the internal region grid is distributed as 130×85×40 leading to total grid size of 1.29 M. Whereas Gird 2 consists of 630×125×50 nodes in outer region and internal region is discretized with 250×150×50 nodes and the total grid is 5.81 M and Grid 3 is 1.5 times that of grid 2. From Figure 2 (Grid 2), it can be seen that the grid clustering is mainly done in the cavity region and also in the aft region to resolve the spatial flow features. The Reynolds number based on the D is $Re_D = 3.39 \times 10^5$. The grid spacing within cavity region for Gird 2 is $\Delta x = 0.02D$, $\Delta y = 0.0067D$ and $\Delta z = 0.034D$, where for both Grid 2 and 3 $y^+$ is 20 and in outer region $y^+$ of 30 is maintained.

At the inlet boundary, uniform flow properties such as static pressure ($P_\infty$ = 18.78 KPa) and static temperature ($T_\infty$ = 107 K) are specified providing Mach 3 flow, while a power law velocity profile is imposed for the velocity. Supersonic freestream condition is imposed at the top, while a no-slip boundary condition is enforced along the bottom wall with condition that the normal pressure gradient vanishes at the wall. At the outlet, flow variables are extrapolated and non-reflecting boundary condition (NRBC) is imposed based on the formulation of Poinsot and Lele [34]. This boundary condition provides a wave transmissive outflow condition based on solving $\frac{\partial}{\partial t}(\psi, U) = 0$ at the boundary. The wave speed is calculated as

$$x_p = \frac{\varphi_p}{|S_f|} + \sqrt{\frac{\gamma}{\psi_p}} \; ;$$

where $x_p$ is the value of the field at the patch, $\varphi_p$ is the flux value of the field; $S_f$ is the face area vector and $\psi_p$ is the patch compressibility. Numerical results are obtained by employing the dynamic SGS model inside the density based solver (rhoCentralFoam) in OpenFOAM framework employing central schemes [35, 36]. These Godunov type central schemes have been used by many authors for various numerical studies [37-39]. The dynamics SGS model utilized for the present computation is based on a dynamic calculation of two model constants. Second order backward Euler scheme is used for temporal discretization while the convection and diffusion terms are discretized using second order low dissipation filtered-linear scheme and central difference scheme, respectively.

## 4. Results and Discussion

This section is organized in the following manner: (1) Initially, the validation of the solver is demonstrated along with the unsteady results, then (2) the results of both energy and enstrophy based POD is discussed at different planes, and (3) finally the observation from the DMD analysis is presented.

**4.1 Mean and Unsteady flowfield**

4.1.1 Grid independence and Mean data analysis

The non-dimensional time averaged pressure distribution for both grids, along the cavity wall is shown in Figure 2. The simulated result (Gird 2 and 3) exhibits excellent agreement with the experimental results of Gruber *et al*. [40]. It can be seen that the variation in pressure is only observed in the region close to trailing edge where the feedback occurs. Since the results obtained via Grid 2 and 3 are in excellent agreement with each other, therefore the detailed analysis is presented in the following sub-sections using Grid 2 only.

To further demonstrate the resolution of the chosen grid, energy spectrum using Grid 2 is presented in Figure 3(a). It can be observed that the spectra follow the -5/3 slope closely in the inertial subrange and hence it can be inferred that the grid is resolved enough to capture the large scale structures in the cavity as well as in the wake region. Apart from energy spectra, index of the grid quality, proposed by the Celik *et al*. [41], which is based on the eddy viscosity ratio, is recast as:

$$LES\_IQ = \frac{1}{1 + \alpha_v \left(\frac{\vartheta_{t,\text{Eff}}}{\vartheta}\right)^n}$$

Where, n = 0.53 and $\alpha_v$ = 0.05. It is generally suggested that LES_IQ above 75 % indicates well-resolved grid for LES computations. In Figure 3(b), LES_IQ for Grid 2 is presented in the near cavity region. It is evitable from the plot that the chosen grid offers sufficient resolution and therefore it is invoked for the detailed analysis of the flow field.

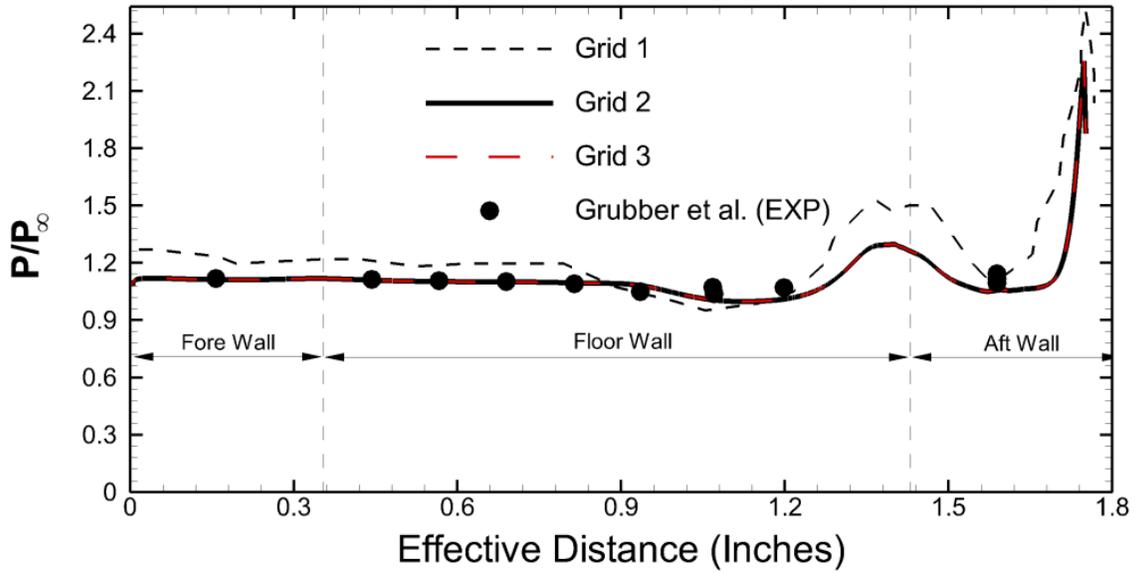

Figure 2: Time-averaged non-dimensional pressure distribution along the cavity wall

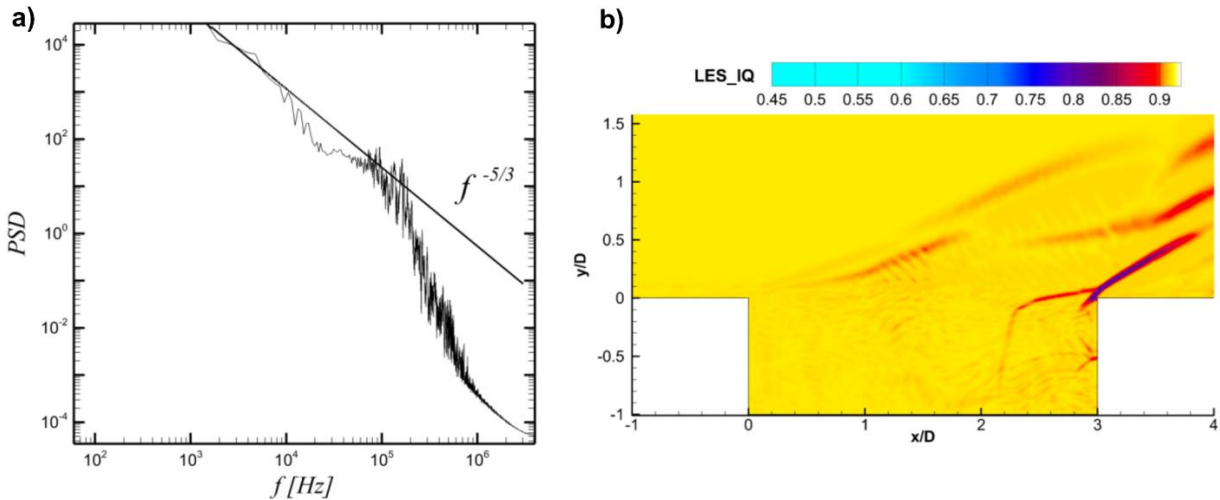

Figure 3(a): Resolved energy spectrum: (b) Resolution of grid through LES quality criteria

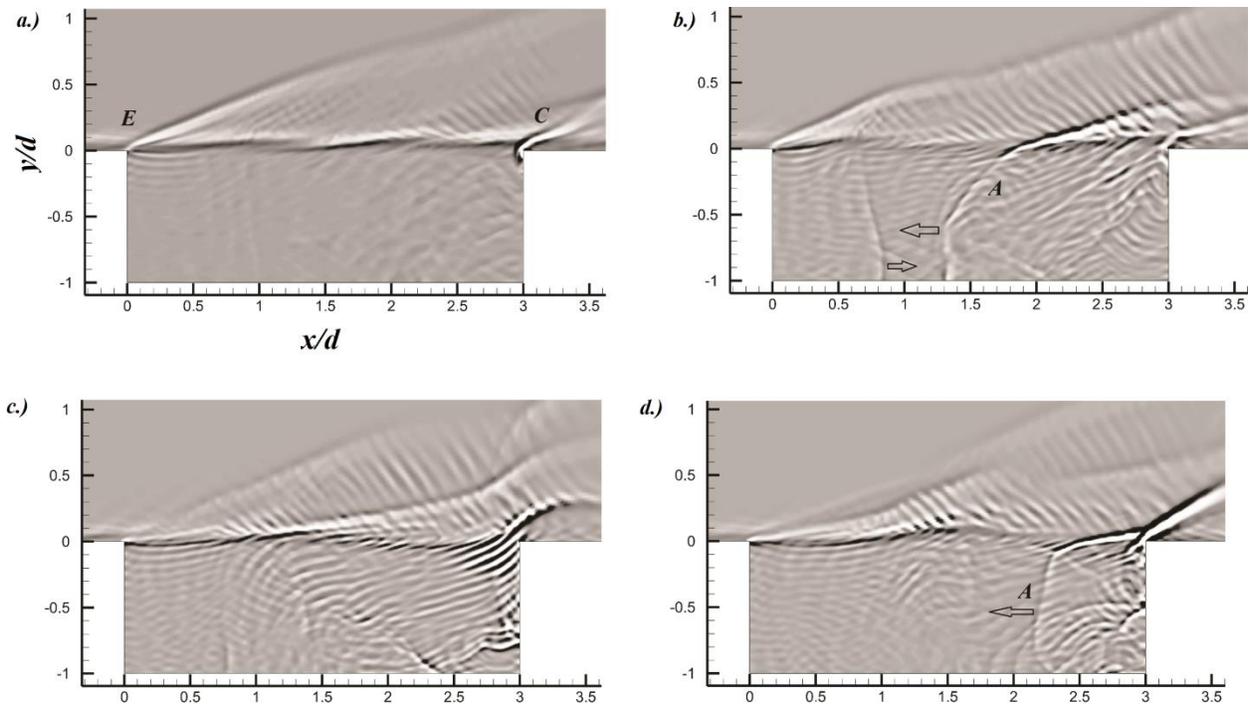

Figure 4: Instantaneous shadowgraph image depicting various flow features a) presence of expansion and compression wave, b) left and right moving acoustic wave c.) Shear layer impingement leading to the formation of acoustic waves d.) Upstream travelling acoustic wave

4.1.2 Instantaneous data analysis

The instantaneous double derivative of density (shadowgraph) is shown in Figure 4. Figure 4(a) corresponds to the mean flow with only expansion (E) and reattachment waves (C) along the cavity leading and trailing edges. In Figure 4(b), apart from the expansion and the compression waves, an acoustic wave (A) traveling left and right can be observed; these acoustic waves while traveling within the cavity perturb the shear layer and lead to self-sustained oscillations. Ben-Yakar and Hanson [42] suggested that the acoustic waves traveling upstream upon impinging the leading edge induce smaller vortices which eventually grow in size as they convect downstream. From the similar figure, it is also evident that the acoustic waves (A) have their leg within the cavity with a part of it moving above the cavity as well. Figures 4(c) and (d) present the formation of compression and expansion wave along the shear layer due to the flapping of the shear layer. The down stroke of the shear layer near the trailing edge introduces the mass inside the cavity which follows the feedback loop and is added back to the main flow with the upstroke of the shear layer. The results presented in the Figures 4 (a) – (d) is consistent with the observation of [43].

Figures 5(a)-(d) present the second invariant of the velocity gradient tensor over one feedback cycle. From Figure 5, it is evident that the large scale structures are convected downstream along the cavity shear layer which is consistent with the observation of [44]. The 3D structures convected downstream upon reaching the trailing edge tend to transform into wake region and are discussed later. Apart from these vortices, there are also vortical structures present near the trailing edge due to vortex shedding. Arya *et al*. [43] also reported K-H vortices along the shear layer, as can be seen, that various three-dimensional structures are present within the cavity and these structures upon interacting with the shear layer induce instability. From Fig 5(a) and Figure 5(b), it is evident that the incoming boundary layer contains vortical structures which transform into spanwise roller right at the leading edge of the cavity.

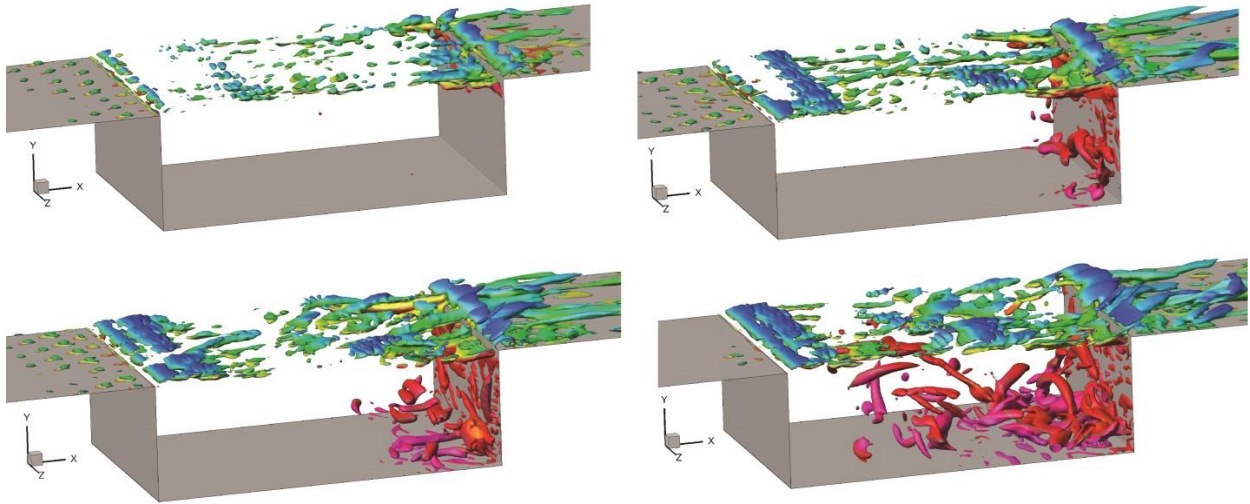

Figure 5: Iso surface of the second invariant of velocity gradient tensor, $Q = 9\left(\frac{u_\infty}{L}\right)^2$ presented for one feedback cycle; coloured by instantaneous streamwise velocity

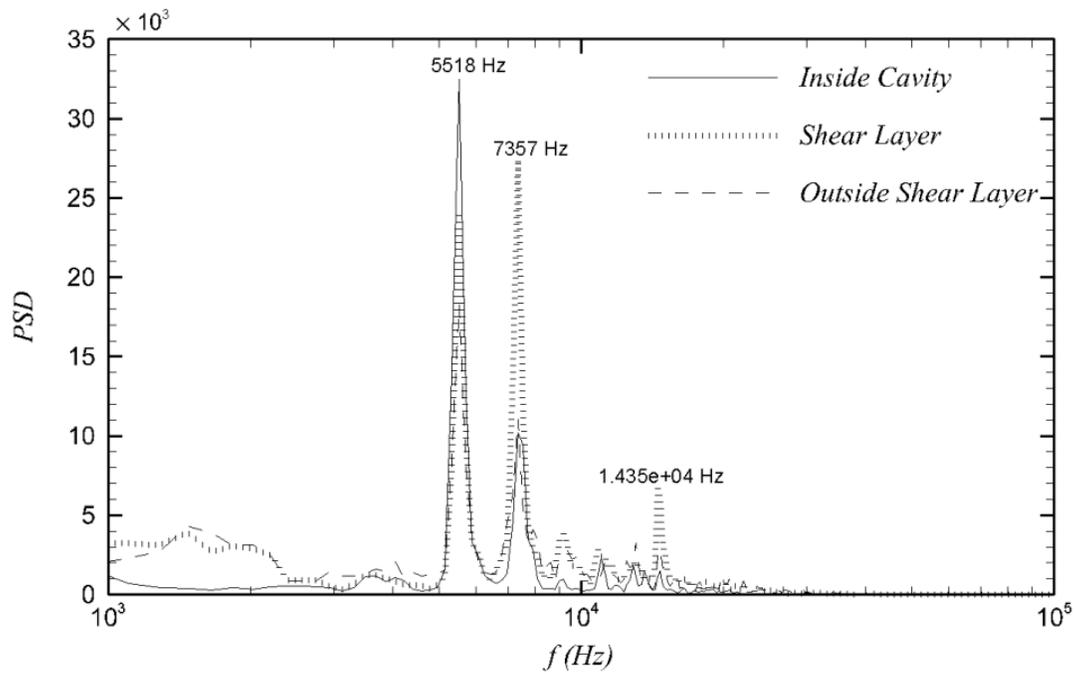

Figure 6: Energy Spectra of pressure signal

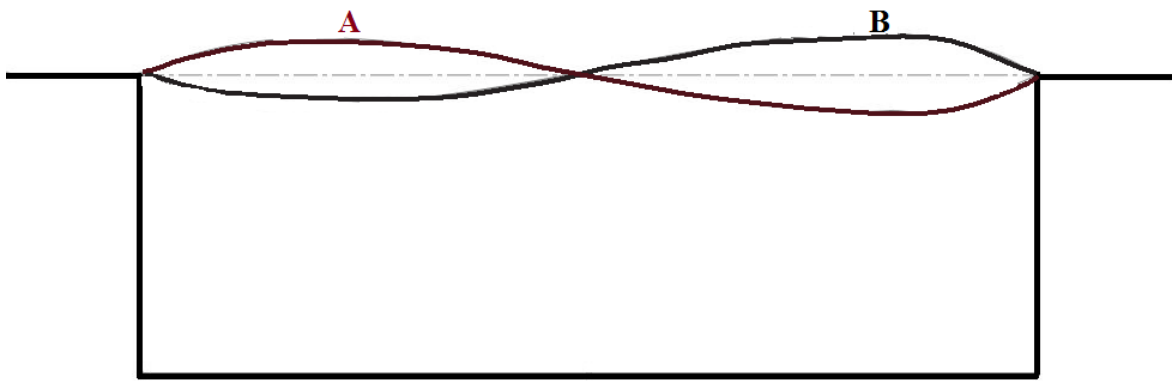

Figure 7: Schematic depiction of shear layer oscillation with mean shear layer position (— · — · —)

4.1.3 FFT and phase-averaging

Figure 6 presents the power spectra where three distinct peaks are clearly visible, while the 2$^{nd}$ and a 3$^{rd}$ peak appear to be harmonic frequencies. To perform FFT, pressure signals are acquired at various locations but the only locations close to the shear layer and trailing edge are presented herein. The first peak corresponds to 5518 Hz and very much close the theoretical calculation of Rossiter [8], where the frequency corresponding to Rossiter's 1$^{st}$ mode calculated through the modified formula is 4834 Hz [8]. To further assess the evolution of flow field within the cavity, phase averaging is performed corresponding to the first two peak frequencies namely 5247 and 7357 Hz. The phase averaging is performed over the 10 cycles with a phase difference of 36º to eliminate the effect of high-frequency oscillations present in the flow field. Figure 7 represents the schematic of shear layer oscillation about the mean shear layer (dotted centerline). The position A of shear layer is representative of upstroke where the mass ejection occurs and position B represents the down stroke leading to mass injection at the trailing edge. The vortices generated at the leading edge induce instability along the shear layer as they are convected downstream and finally impinge on the trailing edge leading to the generation of noise. The vortices that impinge on the trailing edge during the downstroke phase are entrained inside the cavity and roll down along the trailing edge wall through the formation of second recirculation bubble and finally leave the cavity during the upstroke phase.

Figure 8 presents the streamlines over one complete cycle, where the movement of these vortical structures during upstroke and down stroke is clearly observed. Moreover, the shear layer oscillation can be identified along the various phases as the oscillation of shear layer alters the recirculation zones within the cavity. At different phases of the cycle, the motion of two recirculation bubble is observed. In the first half of the cycle (phase 1-5) the leading edge bubble grows in size while the latter half of the cycle (phase 6-10) the trailing edge bubble grows in size. Worth noticing is the significant variation in the evolution of two recirculation bubbles over a cycle, as the core of recirculation zone appears to shift in both transverse and streamwise direction which can be attributed to the shear layer flapping. Essentially, the upstroke of the shear layer results in the mass addition at the trailing edge and the contraction near the trailing edge as the shear layer moves downwards; this phenomenon leads to the change in the size of the secondary bubble. While this leads to the elongation of second recirculation bubble in streamwise direction but during down stroke the reverse happens, means the first bubble grows due to the contraction at the leading edge. This becomes a repetitive process corresponding to the frequency 7357 Hz and primarily happens due to the shear-layer oscillation during upstroke and down stroke. This particular process is more prominent for the other frequency observed in Figure 6 as depicted in Figure 9.

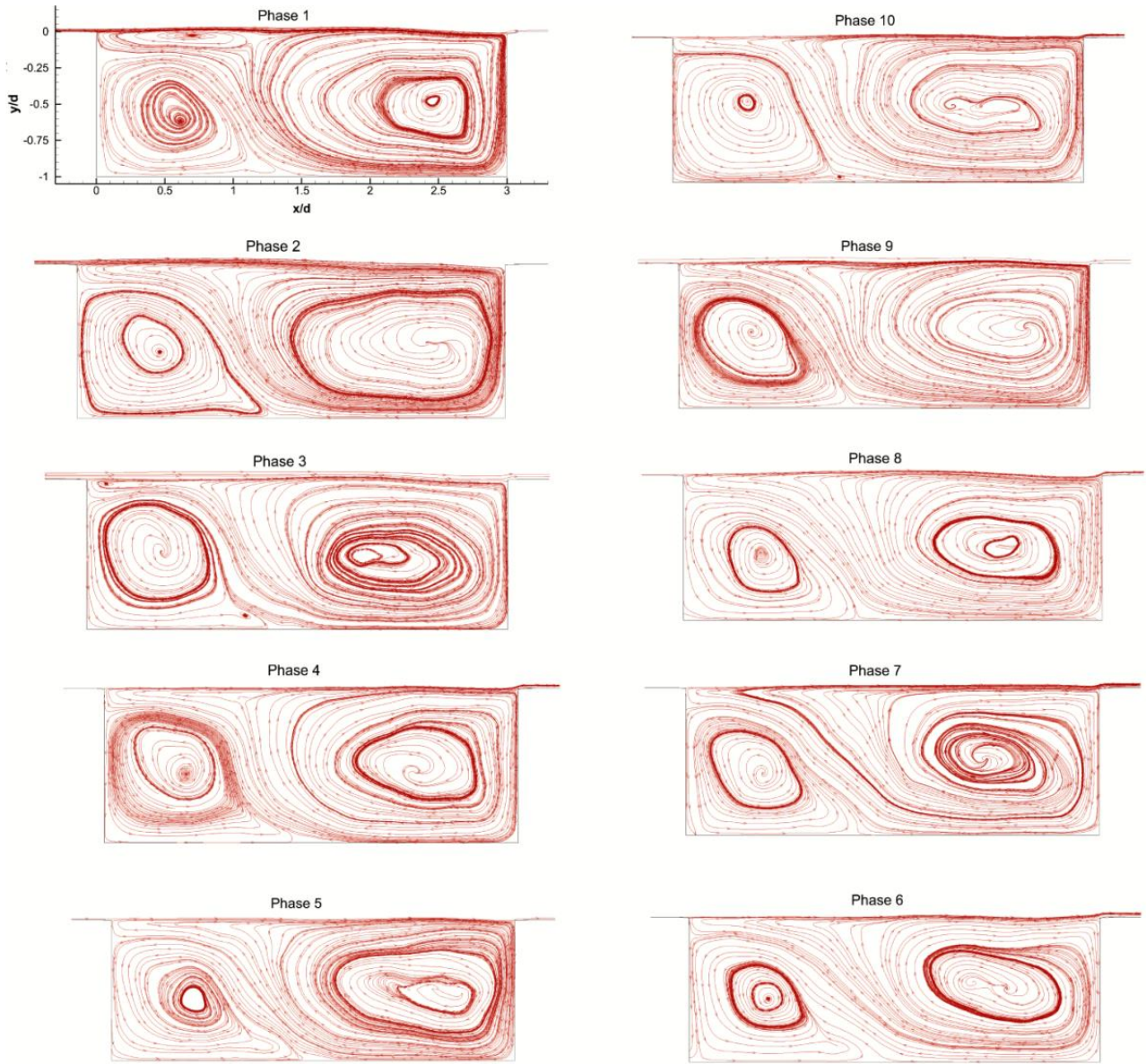

Figure 8: Phase averaged streamline plots over a cycle corresponding to 1$^{st}$ peak ($f = 5247\ Hz$)

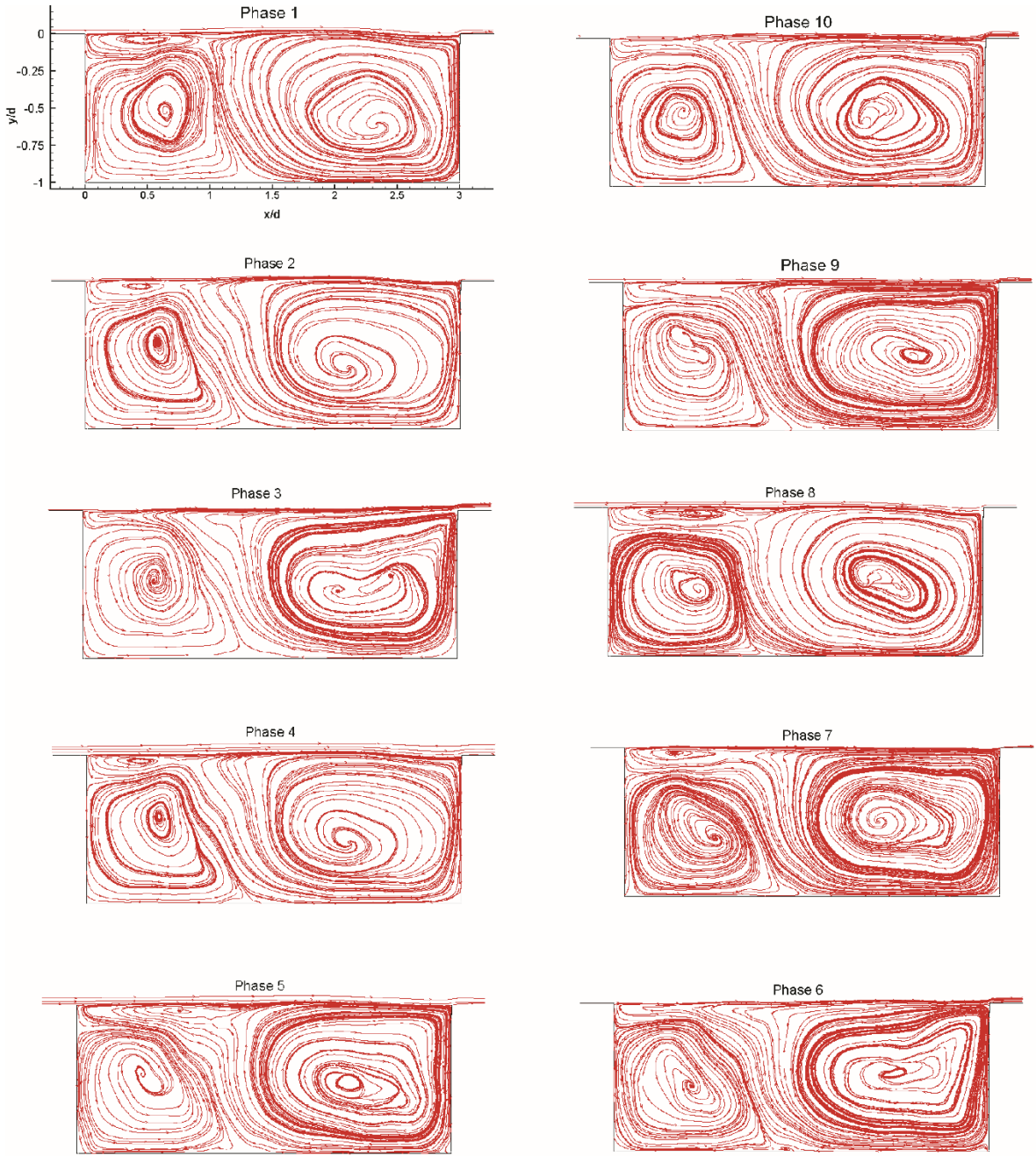

Figure 9: Phase averaged streamline plots over a cycle corresponding to $2^{nd}$ peak ($f = 7357\ Hz$)

In Figure 9 phase averaged data corresponding to the $2^{nd}$ frequency, as obtained from Figure 6, is presented. From different phase plots, the presence of vortical structures close to the cavity leading edge is observed. On combining this observation with that of Figure 5 and Figure 6, it becomes clear that these vortical structures are the spanwise rollers formed at the leading edge which starts pairing at the downstream, clearly visible in phase 8. These large scale vortices (low frequency) which convect over the trailing edge are responsible for the upstream traveling compression waves and the vortices which travel along the shear layer upon impinging the trailing edge wall introduce acoustic excitation. The streamline plots from Figures 8 and 9 clearly demonstrate the presence of the discrete vortices. High degree of unsteadiness in the recirculating bubble is evident from the phase averaged plots. Further details about the underlying physics are discussed in the POD section.

Table 1: Comparison of energy distribution for 1st five Eigen modes for all three N

| Modes | $\lambda_i$ (N = 60) | $\lambda_i$ (N = 80) | $\lambda_i$ (N = 100) |
|---|---|---|---|
| 1st | 0.517 | 0.511 | 0.52 |
| 2nd | 0.136 | 0.123 | 0.13 |
| 3rd | 0.105 | 0.107 | 0.107 |
| 4th | 0.061 | 0.054 | 0.06 |
| 5th | 0.038 | 0.04 | 0.04 |
| 6th | 0.026 | 0.0277 | 0.0277 |

**4.2 POD Analysis (Energy Based)**

4.2.1 POD at $z/h = 0$ plane:

The energy based POD analysis is presented in this section, hereafter energy based POD means that temperature along with components of velocities is used as input to perform the decomposition and the inner product [20-22]. To demonstrate the least number of snapshots required to resolve the coherence in the flow, three different numbers of snapshots (*N*) are initially used and the distribution of energy is demonstrated in Figure 10. For all the values of N presented in Figure 10, the temporal spacing between the consecutive snapshots has been maintained at $\Delta t = 10^{-05}$ s. The energy content of the first mode for all values of *N* is around 51% whereas for the second and third mode for all *N* it varies between 10%-13%. The first 10 modes represent around 88% energy of the mean flow and hence the higher modes which contain energy distribution less than 1% are not included in the further study. Comparison of energy distribution for first five modes for all three *N* is presented in table 1. From Table 1 comparison of eigenvalues for three different values of *N* reveal that there exist very little differences in the value of $\lambda_i$ for a given mode using *N*=80 and 100. Hence in the present section POD, eigenmodes obtained with 80 snapshots are only reported.

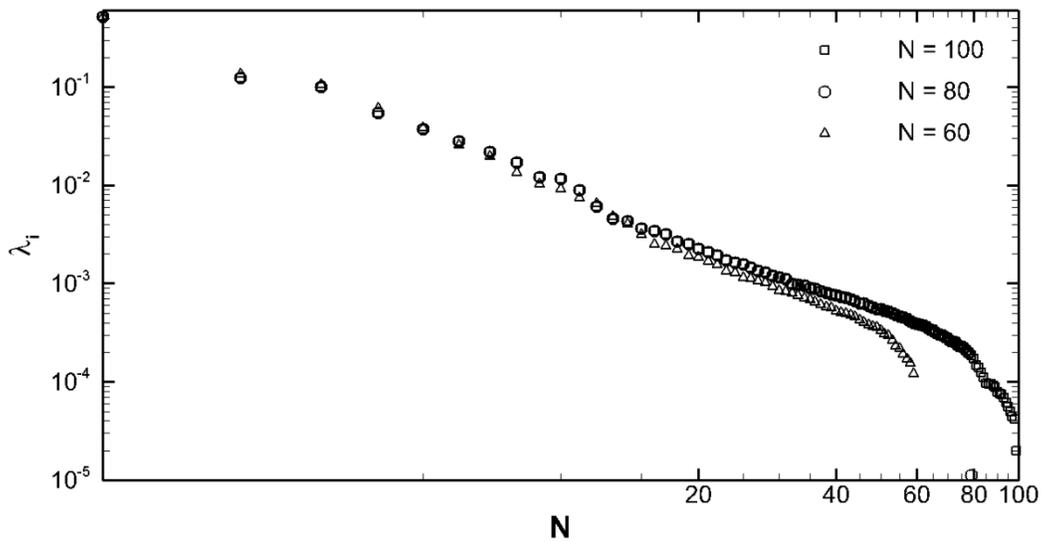

Figure 10: Distribution of relative energy ($\lambda_i$) content as a function of mode number (N)

Figure 11 presents the first six POD modes along the centerline plane (z = 0), the contour represents the streamwise velocity fluctuations. It can be observed that the various coherent structures are revealed by the POD depending on the energy distribution for different modes. The various vortical structures observed here is consistent with the observation of phase averaged and the second invariant of velocity data. The coherent structures are distinguished by the labels, as can be seen from the Figure 11, e.g. type I, type II and so on, to facilitate the discussion. It is evident that every coherent structure exhibits specific characteristics but some appear to transform into different structures altogether. At first, the role and evolution of different vortices are discussed and then the POD modes itself are characterized. Type I structures near trailing edge is formed due to the impingement of discrete vortices present in the shear layer. As discussed in [10] compression and expansion waves are generated at the leading edge, this phenomenon leads to the generation of type II vortices which are convected above the shear layer. Type III vortices are formed due to the upstream traveling disturbances. Further details will be discussed in the following section. Type IV vortices are the structures present inside the boundary layer, whereas type V structures are originally part of type I vortices which follow the feedback loop and type VI vortices are formed by the pairing action of type II vortices.

In mode 1 type I and II vortices are resolved, the type I vortices upon impinging the trailing edge leads to the acoustic resonance and an upstream traveling disturbance is created which interacts with the shear layer at the leading edge. The type II vortices are present above the shear layer and can be associated with the large-scale structures that are responsible for the wake formation at the cavity trailing edge. From mode 2 and 3, a very important observation can be made; type III vortices, which are also observed in the Figures 8 and 9, is generated at the leading edge (mode 2) and convect along the shear layer (mode 3). From the discussion in previous sections, one can figure out that these vortices are formed due to the perturbation caused by the shear layer. The addition of mass within the cavity at the trailing edge leads to the perturbation at the leading edge and hence type III vortices which are basically low frequency structures are generated. The presence of the type III structure is consistent with the observation of Figure 8 and 9 and confirms the presence of the feedback loop which suggests that the present modal decomposition is able to resolve the structures formed due to both hydrodynamic as well as acoustic modes accurately. Apart from these, the low frequency structures of type I are present along the trailing edge the two structures appear to be of similar wave length. On combining this observation with that of Figure 5 and 6 impinging it would become apparent that actually it the type I vortices which are transformed into type III structures due to pairing/merging while convecting upstream. The presence of the vortices in the boundary layer is noticed, it is possible that while convecting downstream they interact with other vortices especially ones formed on the outer edge of the shear layer. The type II vortices present on either side of the shear layer are present across different modes. These vortices start to pair close to the trailing edge and the pairing continues further downstream leading to the large scale structure in the wake region (type VI). The overall convincing output is obtained through the modal decomposition, i.e., all the vortical structures corresponding to major phenomena are resolved. Although both hydrodynamic and acoustic mode related structures are present but the discussion in a temporal sense may be not be convincing from the eigenmodes plot alone. However to remedy this combined analysis, i.e. instantaneous data set can be utilized to better understand the nature and formation of certain vortical structure and their motion.

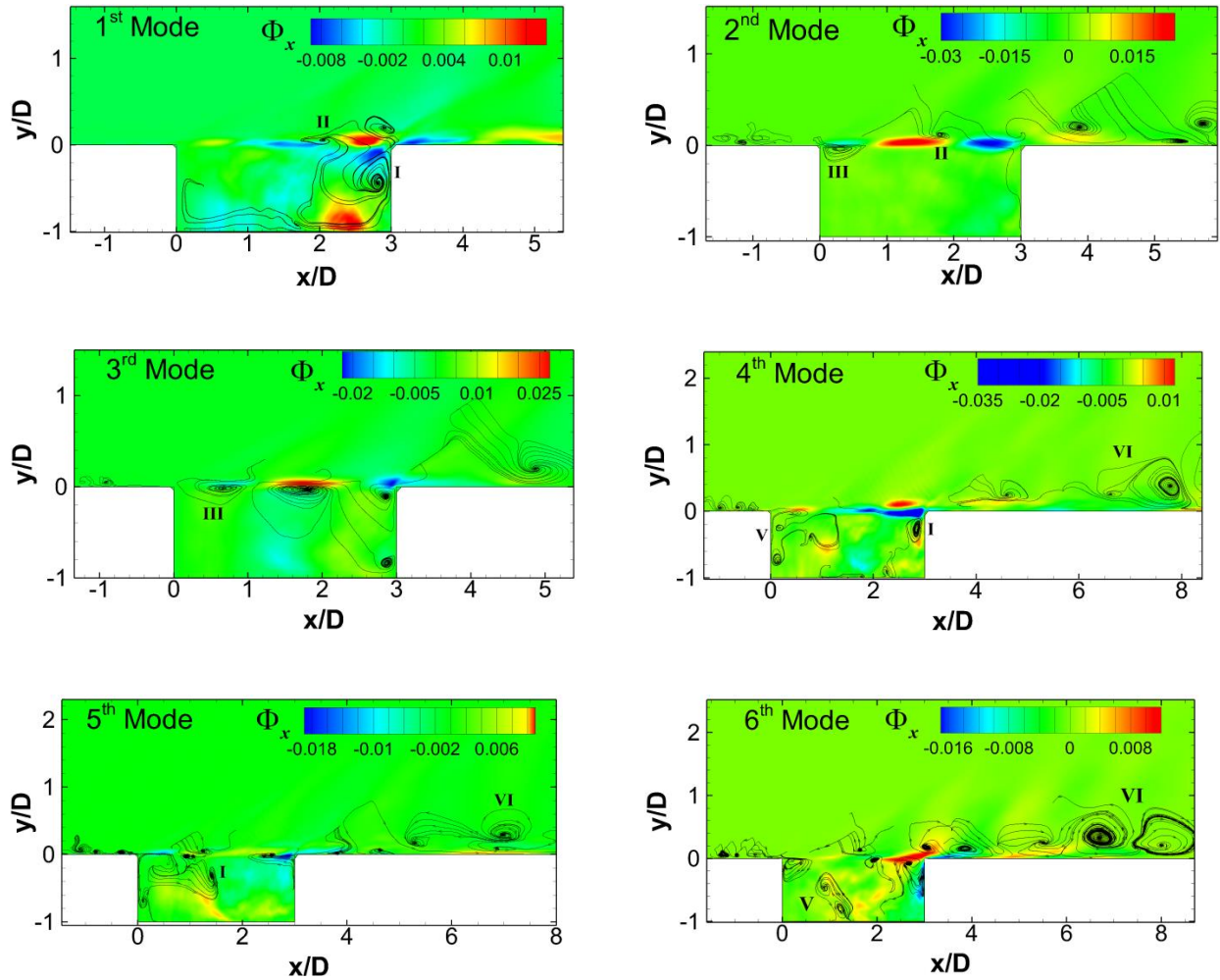

Figure 11: First six spatial eigenmodes colored with the streamwise velocity fluctuation

Now coming onto the contour plot it can be noticed that all 6 modes exhibit two local extremes of opposite sign. This observation is especially more pronounced in the first 4 modes, modes 1 and 4 appear to be identical; similarly, modes 2 and 3 appear in a pair but with the spatial shift compared to modes 1 and 4. This suggests the periodicity in the presence of the large scale structures, present along the shear layer. The first local extreme presents in $1^{st}$ and $4^{th}$ mode are of positive sign and suggest the local increase of streamwise component yielding the downward shift of the shear layer, whereas the second local extreme of negative sign induces the increment in the streamwise component and results in upward movement of the shear layer. This can be ascribed to the momentum ejection and injection across the shear layer. On combining this argument with the first 4 modes, it becomes evident that the shear layer flapping is a periodic event oscillating regularly about the mean position. This suggests that the first 4 modes represent the undulating motion of the shear layer. The last two modes ($5^{th}$ and $6^{th}$) are mainly representative of wake mode and relatively small-scale motions.

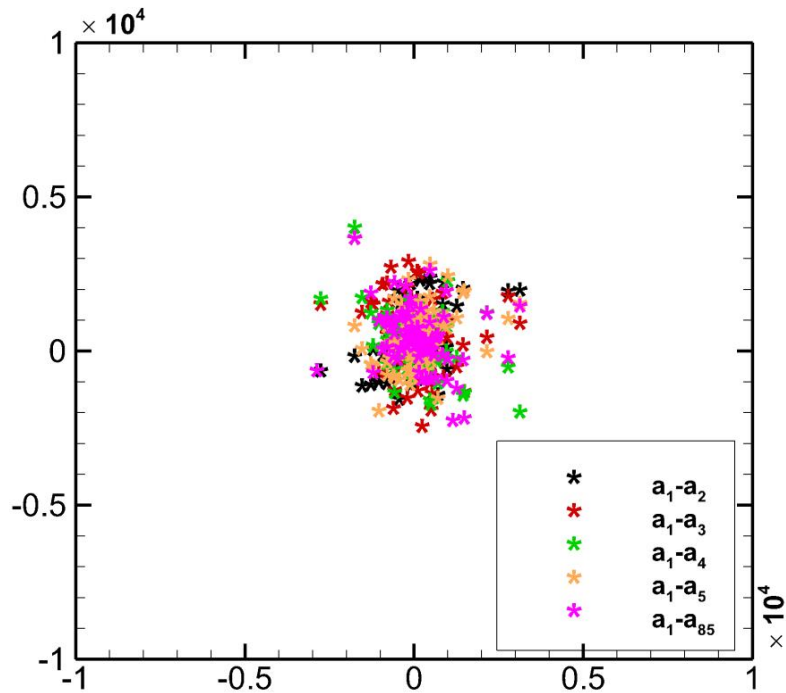

Figure 12: Phase plots of temporal coefficient

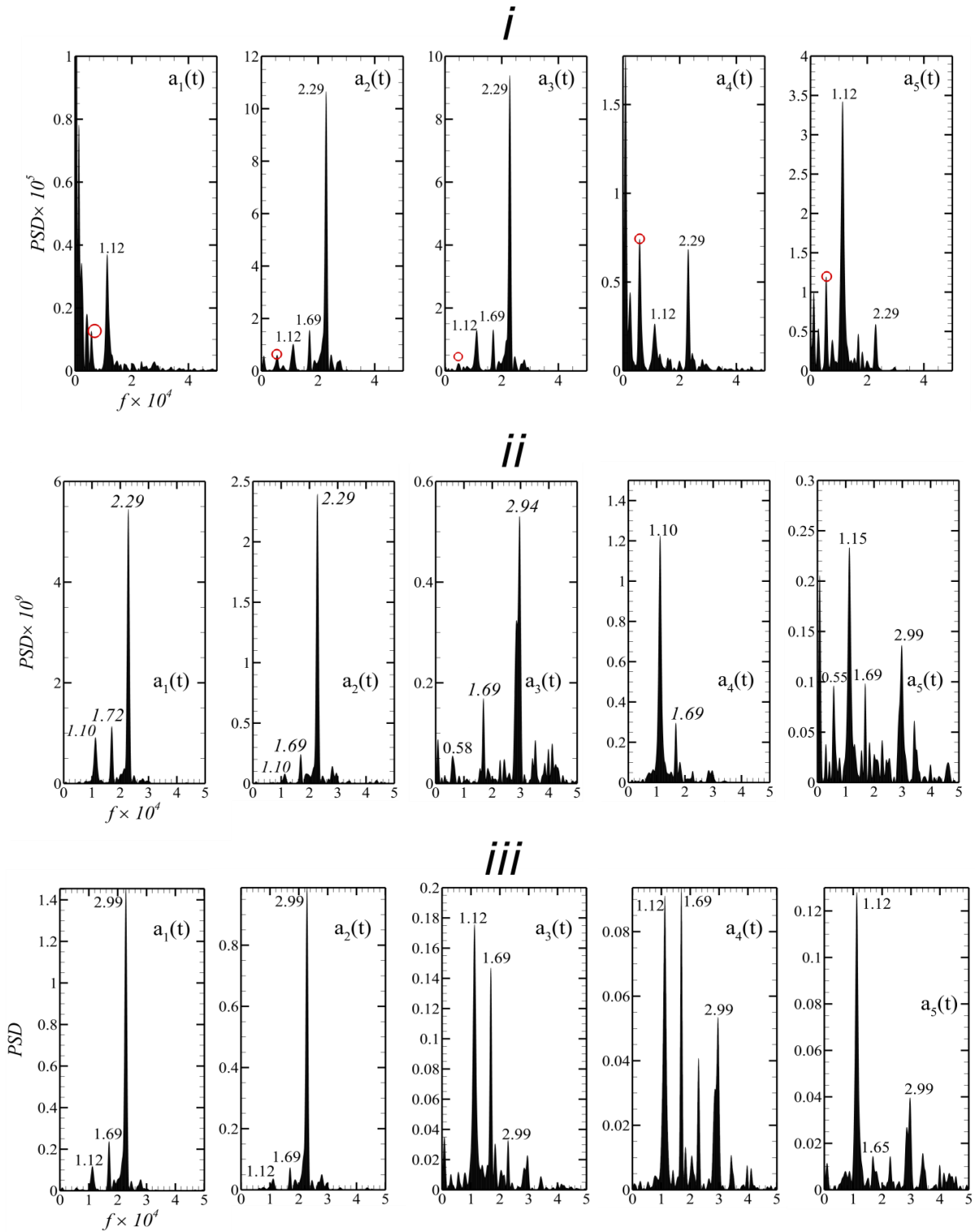

Figure 13: Fast Fourier transform of the temporal coefficient corresponding to the 1st five most energetic modes (i. Energy POD, ii. Pressure POD and iii. Density POD)

The phase plots of temporal coefficients are presented in Figure 12. The presence of a low dimensional attractor is clearly visible from the plots as mostly temporal coefficients are clustered around the origin. Hence, it can be inferred that the decomposition does its job to perform a linear mapping of the data to a lower dimensional space such that the variance of the data in the lower dimensional space is maximized.

The eigenmodes presented in the Figure 11 put forward the spatial coherence in the flow field but does not provide information about the temporal coherence although POD is capable of revealing temporal coherence but still some information can be deduced. Since the decomposition works by separating the flow field in time and space eigenmodes provide information about the spatial coherence, however, the temporal coefficients can be invoked to get insight in the spectral sense. But one must be careful while carrying out such exercise as POD does not reveal the pure frequency instead of being evaluated based on the energy content across modes contains various frequencies. Hence the FFT of the first five temporal coefficients for three different norms is evaluated and collated in Figure 13. Since the spatial coherence is established from the energy based eigenmodes, hence for the sake of brevity only FFT of temporal coefficients for pressure and density based decomposition are reported. The peak highlighted with the red circle corresponds to 5247 Hz which corresponds to the vortical motion along the shear layer (Figure 8). This explains the presence of this particular frequency among all the eigenmodes investigated in Figure 11. It can be thought of as the most probable (or dominant) frequency present in the domain, one can safely assume that this frequency is responsible for large scale vortical motion along the shear layer and hence is the frequency of the shear layer oscillation as well. The relation between mode 2 and 3 in Fig 13 i.) is clearly unfurled from the FFT as well, the mode seems to be phase shifted suggesting convection of vortices along the shear layer. Similarly, for the density and pressure based decomposition mostly common frequencies are observed. However, for these two decompositions mostly higher harmonics is witnessed which suggests that these decompositions represent the acoustics related phenomenon. Most intriguing is the presence of the higher frequencies; it appears that the two peaks are in fact super harmonic of the shear layer flapping frequency. The harmonics can be related to the small scale vortical motion happening within the cavity due to vortex stretching/breakdown at the trailing edge.

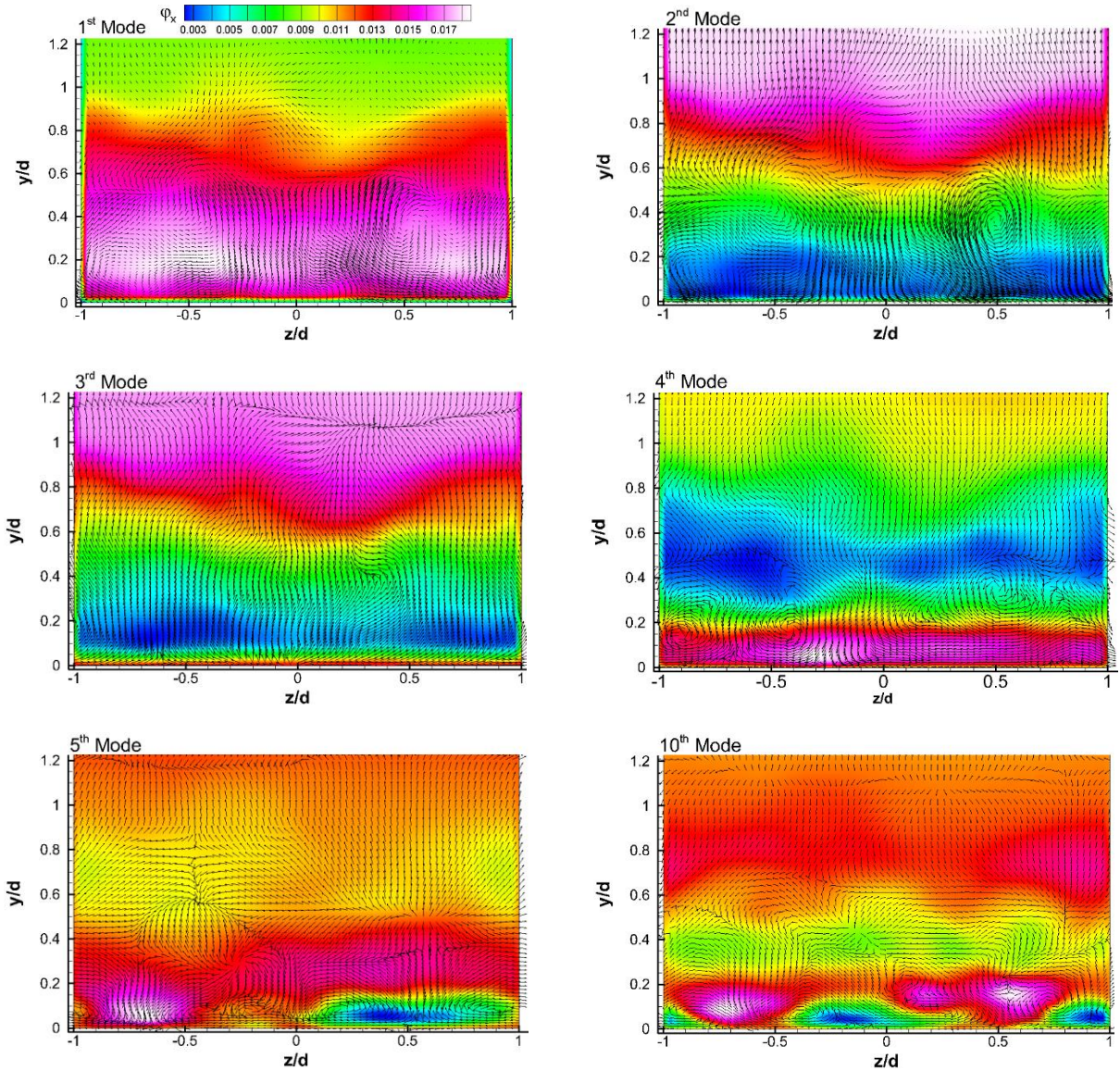

Figure 13: POD modes for $x/d = 3.93$ plane (Contour: Out of –plane component)

4.2.2 POD at $x/d = 3.93$ plane:

POD modes for x/d =3.93 plane are presented in this section which lies in the region of vortex shedding immediately after the reattachment point. Here again, POD results are shown for the N=80 case. Again first 5 and 10$^{th}$ eigenmodes are presented in this section. From Figure 13 it can be inferred that 1$^{st}$ and 3$^{rd}$ mode is closely related with 1$^{st}$ mode very close to mean flow. In these two modes, vortical structures are observed above the aft region wall whereas 2$^{nd}$ and 4$^{th}$ modes, which again appear similar, have more pronounced vortical structures shifted toward the wall region. Similarly, from 5$^{th}$ and 10$^{th}$ mode it can be seen that the structures are clustered close to the wall. On combining this observation with that of z/d = 0 plane, it can be concluded that indeed large scale structures are present due to the vortex shedding at the trailing edge as well as the structures passing through the shear layer, and this is consistent with the observation of Figure 11.

## 4.3 DMD Analysis

From the discussion of the proper orthogonal decomposition in the preceding section the spatial coherence is established. Although the fast Fourier transform does offer significant insight in the spectral sense, presence of mixed frequencies throughout the modes indeed creates the confusion. The source of this confusion lies in the lack of POD to extract temporal orthogonality which is its inherent limitation. This issue can be remedied by invoking another modal decomposition technique that extracts the temporal orthogonality and relieves the observation of the mixed frequency. The dynamic mode decomposition does not collate the modes on the basis of energy content instead it is based on the observed frequencies which are exclusive to a particular mode. So in a way it can be said that the eigenvalues associated with the DMD are actually representative of the growth and decay of any instability present in the flow field. In the present investigation $N$= 101 snapshots are utilized which results in the companion matrix of the size *(N-1) × (N-1)*. The snapshots for the DMD has been generated such that it complies with the Nyquist criteria, in the current investigation the sampling frequency ($F_s$) has been maintained to be 10$^5$ Hz. This choice of the $F_s$ will assure that all the frequencies of the interest are resolved especially those detected in the POD temporal coefficient spectra.

In Figure 14 $L_2$ - norm of the DMD spectrum for pressure and velocity are presented. The choice of the sampling frequency is reflected in the resolution of all the frequencies that are relevant to characterize the physics. The three frequencies marked in the plot (Figure 14) as 1, 2 and 3 represents 5590 Hz, 7381 Hz and 11190 Hz, respectively where the highest frequency is the super harmonic of the lowest one. The dynamic modes corresponding to all the three field variables are presented for only those three frequencies, which is consistent with the observation of the temporal coefficient spectra. In Figure 15 a.) –c.) the dynamic modes computed for velocity, vorticity and pressure respectively, are presented. The dynamic modes related to streamwise velocity show the convection of large scale vortical motion along the shear layer and inside the cavity. The structures present along the shear layer are the ones generated at the leading edge and convecting downstream to interact with the trailing edge and lead to the initiation of the self-sustained oscillation (feed-back mechanism). The large wavelength present inside the cavity for first two modes signifies the low frequency structures traversing upstream whereas for the last mode relatively smaller wavelength structure inside the cavity giving impression of the feedback cycle is present. It can be conjectured that $1^{st}$ two mode reflect mostly the large scale vortices whereas the $3^{rd}$ mode signifies the effect of vortex stretching in-side the cavity while sweeping upstream. Apart from these, structures past trailing edge is also reflected in the modal plot which suggests that shear layer flapping and vortex shedding in the wake region are at the similar frequency. The same is true for the dynamic mode computed through spanwise vorticity; for low frequency, large wavelength are present within the cavity and for the super harmonic, smaller wavelength depicting a complete feed-back cycle is present. The presence of the local maxima and minima characterizes the vortical motion of opposite sign convecting downstream. The large wavelength witnessed for the lower frequencies probably signifies the interaction of vortex leading to pairing/merging phenomenon. The presence of single structure along the shear layer is possibly related to the Rossiter's $1^{st}$ mode which makes sense because as seen from the POD analysis low frequency type III vortices are present due to the acoustic perturbation. However, the single structures for highest frequency showing two wave lengths along the shear layer signifies the vortex stretching process possibly higher Rossiter's mode. Finally, the modes corresponding to the pressure mostly represent the propagation of acoustic wave inside the cavity. The acoustic wave travels back and forth inside the cavity and upon interacting with the shear layer at the leading edge results in the formation of vortices in shear region leading to the onset of the self-sustained oscillation. The different pressure modes also seem to represent the compression and expansion region at the leading and trailing edge apart from the waves that interact with the shear layer during up and down stroke. For example the first two low frequency modes clearly show the presence of bow shock (or reattachment) at the trailing edge as region of high gradient.

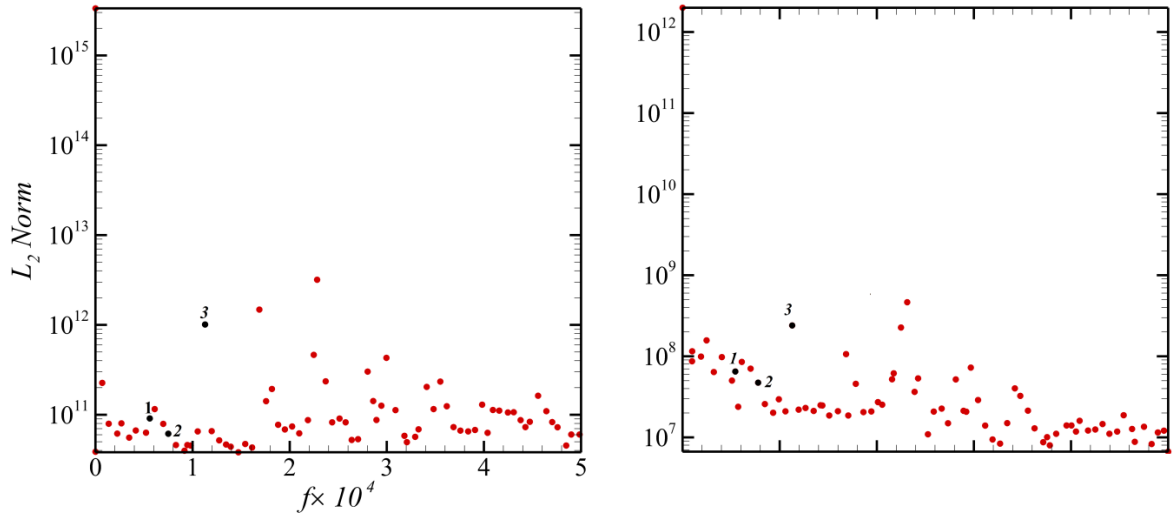

Figure 14: $L_2$ norm for pressure (left) and streamwise velocity (right)

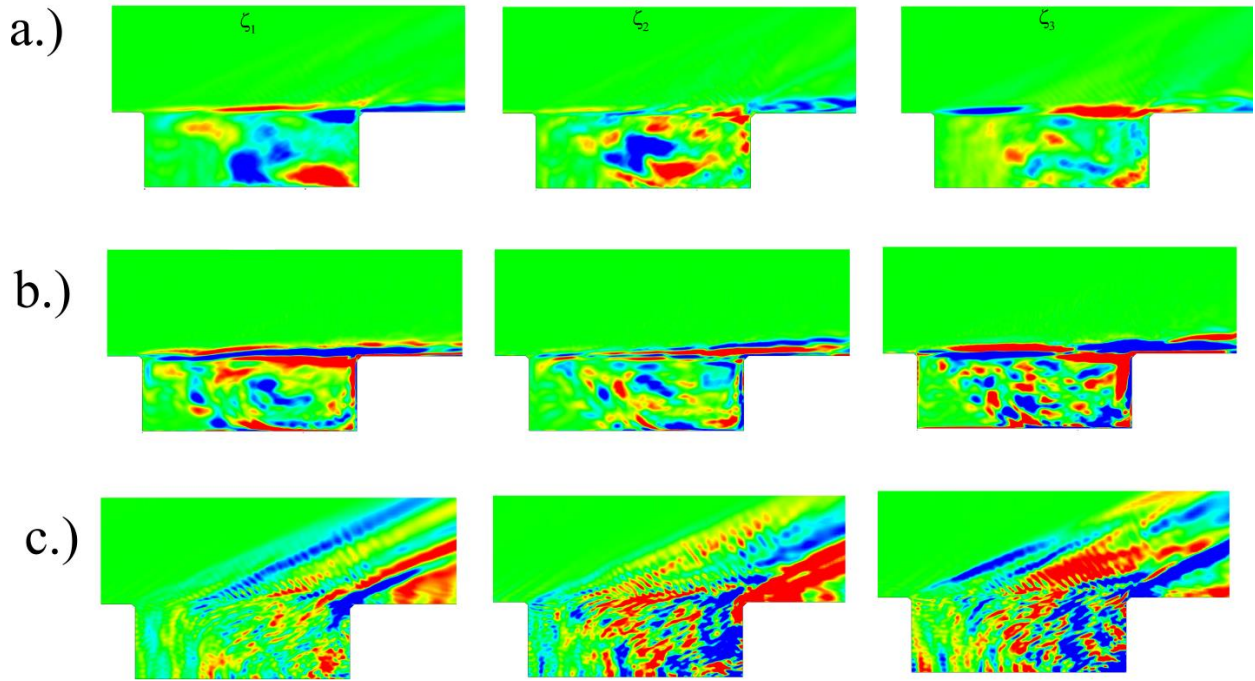

Figure 15: Dynamic modes computed for a.) Streamwise velocity, b.) Spanwise vorticity and c.) Pressure field

## 5. Conclusions

Wall modeled LES of the supersonic open cavity at Mach 3 is reported, and the numerical results (mean) follow the experimental observation very closely. The numerical results validate the fact that wall modeling approach is efficient in reducing computational overhead without compromising the numerical accuracy. Also, the LES_IQ demonstrates that the effect of grid resolution used in the present study. Unsteady data and phase-averaged data reveal the presence of discrete vortices along the shear layer, it is also confirmed that these vortices are pretty much responsible for the acoustic excitation and feedback mechanism in the cavity. The modal decomposition (POD and DMD) is performed on the various snapshots and it reveals that major flow structures governing much of the phenomenon in the cavity are discrete vortices present along the shear layer. The energy and scalar based POD is performed along the centerline plane ($z = 0$). The energy based POD along this plane reveals the evolution of various large scale structures. Higher eigenmodes along this plane reveal the presence of discrete structures formed due to the acoustic excitation and wake vortices; also incoming boundary layer contains vortical structures which are convected downstream. The variation in the wavelength across various modes suggests phenomena exclusive to vortical motion, i.e. stretching and pairing/merging. This suggests that the vortices growth is basically due to the interaction of different type of vortices generated independently. The spectra of temporal coefficients for energy, pressure and density show consistency in a sense that the entire norm reveals more or less similar frequencies. The fundamental frequency that is close to the Rossiter's first mode is also revealed most prominent being the $1^{st}$ super harmonic which is present consistently throughout for all the norms. The DMD performed on the spanwise vorticity, streamwise velocity and pressure all reveal the similar observation as POD. Presence of the feedback mechanism is established through the velocity and vorticity based decomposition whereas pressure based modal analysis points toward the acoustics related phenomenon.

**Acknowledgments**

Financial support for this research is provided through IITK-Space Technology Cell (STC). Also, the authors would like to acknowledge the High-Performance Computing (HPC) Facility at IIT Kanpur (www.iitk.ac.in/cc).